\documentclass[%
 reprint,
superscriptaddress,
 amsmath,amssymb,
 aps,
]{revtex4-2}

\usepackage{graphicx}% Include figure files
\usepackage{dcolumn}% Align table columns on decimal point
\usepackage{bm}% bold math
\usepackage[version=4]{mhchem}
\usepackage{xcolor}
\usepackage{booktabs}
\usepackage[normalem]{ulem}
\newcommand{\Eqref}[1]{Equation~\eqref{#1}}
\newcommand{\Figref}[1]{Figure~\ref{#1}}
\newcommand{\Tabref}[1]{Table~\ref{#1}}

\newcommand{\mnew}[1]{\textcolor{black}{#1}}
\newcommand{\mrevrev}[1]{\textcolor{black}{#1}}

\newcommand{\bergen}{Department of Physics and Technology, University of Bergen, Allégaten 55, 5007 Bergen, Norway}
\newcommand{\pisa}{Center for Nanotechnology Innovation@NEST, Istituto Italiano di Tecnologia, Piazza San Silvestro 12, 56127 Pisa, Italy}
\newcommand{\donostia}{Donostia International Physics Center (DIPC), Paseo Manuel de Lardizabal 4, 20018 Donostia-San Sebastián, Spain}

\begin{document}

%\preprint{APS/123-QED}

\title{
Observation of increasing bending rigidity of graphene with temperature 
}

\author{Martin T\o{}mterud}
\email{martin.tomterud@uib.no}
\affiliation{\bergen}

\author{Simen K. Hellner}%
\affiliation{\bergen}

\author{Sabrina D. Eder}%
\affiliation{\bergen}

\author{Stiven Forti}%
\affiliation{\pisa}

\author{Domenica Convertino}
\affiliation{\pisa}

\author{Joseph R. Manson}
\affiliation{\donostia}%
\affiliation{Department of Physics and Astronomy, Clemson University, Clemson, South Carolina 29634, USA}%

\author{Camilla Coletti}%
\affiliation{\pisa}

\author{Thomas Frederiksen}
\affiliation{\donostia}%
\affiliation{Ikerbasque, Basque Foundation for Science, 48013 Bilbao, Spain}%

\author{Bodil Holst}
\email{bodil.holst@uib.no}
\affiliation{\bergen}

\date{\today}

\begin{abstract}
The mechanical properties of two-dimensional materials are important for a wide range of applications including composite and van der Waals-materials, flexible electronics and superconductivity. Several aspects are highly debated in the literature: For example, the theoretically predicted bending rigidity $\kappa$ at 0 K for quasi free-standing graphene varies from 0.8 to 1.6~eV, and there are predictions that it could either increase or decrease with temperature. Here we present an experimental study of the temperature-dependent bending rigidity $\kappa(T)$ of graphene. From the phonon dispersion relation measured with helium atom scattering for the out-of-plane acoustic (ZA) mode, we find $\kappa(T)$ to increase with sample temperature. We compare our experimental results with novel molecular dynamics (MD) simulations performed as part of this study as well as available literature data. The calculations reproduce the temperature trend of our experiments, but with a slightly weaker slope. A probable cause for the observed differences is the slight strain associated with experimental substrate supported graphene that is not present in the calculations.
\end{abstract}

\maketitle

\section{Introduction}\label{sec1}

The ever expanding family of two-dimensional (2D) materials is continuously showing promise for numerous applications, mainly due to the extraordinary properties associated with them. Among these characteristics are great mechanical strength~\cite{Papageorgiou2017}, tunable electrical properties and high electrical conductivity~\cite{Novoselov2004}, favourable properties for use in electronic and optoelectronic devices~\cite{Wang2012, Jariwala2013, Chaves2020}, and  superconductivity~\cite{Cao2018}. 2D materials figure as building blocks in layered structures, commonly known as van der Waals heterostructures~\cite{Liu2016}. The materials are promising candidates for use in flexible electronics and bendable electronic devices~\cite{Akinwande2014}. 

The flexibility of a material is expressed through the \textit{bending rigidity}, which is also referred to as the flexural rigidity or the bending stiffness. Classically, the bending rigidity $\kappa$ of an isotropic plate is given by~\cite{Landau2012}
\begin{equation}
    \kappa(h)=\frac{Y h^3}{12(1-\sigma)},
    \label{eq:one}
\end{equation}
where $Y$ is Young’s modulus, $h$ the plate thickness, and $\sigma$ Poisson’s ratio. For hexagonal crystals, such as graphene, the elastic properties in the (0001) plane can be treated as isotropic~\cite{Landau2012}. Evaluating the bending rigidity using \Eqref{eq:one} has caused debate in the graphene community as there have been several measurements of Young's modulus with a wide range of values reported~\cite{Wang2005,Pine2014,Huang2006}. Furthermore, there is no clear indication that the classical result of \Eqref{eq:one} should hold for atomically thin sheets. In particular, if one assumes that the thickness of bilayer graphene is twice that of monolayer graphene, \Eqref{eq:one} predicts that $\kappa$ of bilayer graphene should be eight times that of monolayer graphene. Simulations have however indicated that $\kappa$ of bilayer graphene only doubles~\cite{Zakharchenko2010} compared to monolayer graphene.

Various techniques of nanoengineering have been applied to measure the mechanical properties of 2D materials~\cite{Akinwande2017}, resulting in a range of values for $\kappa$ of monolayer graphene at room temperature. 
Zhao \textit{et al.} reported $\kappa \in 1.8-2.7$ eV using high-resolution STEM imaging~\cite{Zhao2015}. Han \textit{et al.} obtained two values, $\kappa = 1.2 \pm 0.11$ eV and $\kappa = 1.7 \pm 0.5$ eV with electron microscopy experiments~\cite{Han2020} by draping graphene over hexagonal boron nitride (h-\ce{BN}) steps of different heights. Blees \textit{et al.} obtained an effective bending rigidity of $\approx 4000 \times 1.2$~eV~\cite{Blees2015}, extracted from graphene that had undergone kirigami processes. An often cited value in the literature  $\kappa = 1.2$~eV, was derived from Raman scattering measurements on graphite by Nicklow \textit{et al.}~\cite{NicklowPRB}. 

Theoretical predictions at 0 K using density functional theory (DFT) have given values from about 0.8~eV to 1.6~eV for monolayer graphene~\cite{Shen2012, Guo2016, Wei2013, Chen2015, Koskinen2010, Zakharchenko2010, Zhang2011, Sanchez-Portal-PRB, Kudin-PRB, Munoz-Diamond, Tersoff-PRB, Tu-PRB, Arroyo-PRB, Lu-APD}. 
Temperature-dependent calculations have suggested both increasing and decreasing values of the bending rigidity as a function of temperature: Atomistic Monte Carlo simulations~\cite{Zakharchenko2010}, simulations of thermal ripples~\cite{Costamagna-PRB}, molecular dynamics (MD) simulations~\cite{Ramirez2016}, and energy variation analysis of carbon nanotubes~\cite{Yi-Aip} all predict that the bending rigidity should increase with temperature $T$, while membrane theory predicts that it should decrease~\cite{Liu-APL}. Very recently, a novel approach for calculating bending and twisting rigidities of 2D materials was proposed by means of a torus-based surface parameterization~\cite{Vel2024}. \mnew{Both theoretical and experimental evaluation of the elastic properties of graphene is possibly made more challenging due to strain-related effects such as buckling and ripples in free standing and supported graphene films~\cite{PhysRevB.86.035427, Dmitriev2012}.}

Amorim and Guinea showed that $\kappa$ for an isotropic, free-standing thin membrane which is weakly bound to a substrate is related to the phonon dispersion of the out-of-plane acoustic (ZA) mode via~\cite{Amorim2013}
\begin{eqnarray}
\omega^2_{\text{ZA}}(q)=\frac{\kappa}{\rho_{\text{2D}}}q^4 + {\omega_0}^2,
\label{eq:two}
\end{eqnarray}
where $\omega$ is the phonon frequency, $\rho_{\text{2D}}$ the 2D material density, $q$ the wave vector, and $\omega_0$ the coupling frequency between the membrane and the substrate.
\Eqref{eq:two} indicates that $\kappa$ is the bending rigidity of the free-standing material. That is to say, as long as the 2D material is only weakly bound to the substrate, the intrinsic bending rigidity can measured for a quasi-free-standing 2D film. \mnew{For an infinitely thin film with fixed-edge boundary conditions, \Eqref{eq:two} should contain also a quadratic term in $q$, c.f.~erratum of Ref.~\cite{Buchner2018}}. This implies a linear term in the dispersion relation,
\begin{equation}
    \omega^2_{\text{ZA}}(q)=\frac{\kappa}{\rho_{\text{2D}}}q^4 + v_{\text{ZA}}^2q^2 + \omega_0^2.
    \label{eq:za_lin}
\end{equation}
where the sound velocity $v_{\text{ZA}} = \sqrt{c_{\text{44}}/ \rho_{\text{2D}}}$ is related to the shear force constant $c_{44}$ of the material. 
However, as discussed in Ref.~\cite{Buchner2018}, this extra term is typically shadowed by the coupling energy and therefore ignored.

Due to the result by Amorim and Guinea, helium atom scattering (HAS) has been established as a technique uniquely positioned for evaluating the bending rigidity of 2D materials~\cite{Holst2021}. This is mainly due to the non-penetrating nature of the HAS technique that allows for measurements of the dynamics of 2D materials in the meV energy range~\cite{Holst2013}. Al Taleb \textit{et al.} applied the theory in 2015 to measure the bending rigidity of free-standing monolayer graphene using phonon dispersion curves from monolayer graphene weakly bound on a copper substrate~\cite{AlTaleb2015}. Their measurement was conducted at 120 K, and they obtained a value of $\kappa = 1.30 \pm 0.15$ eV. 
\begin{figure}
    \centering
    \includegraphics[width = \columnwidth]{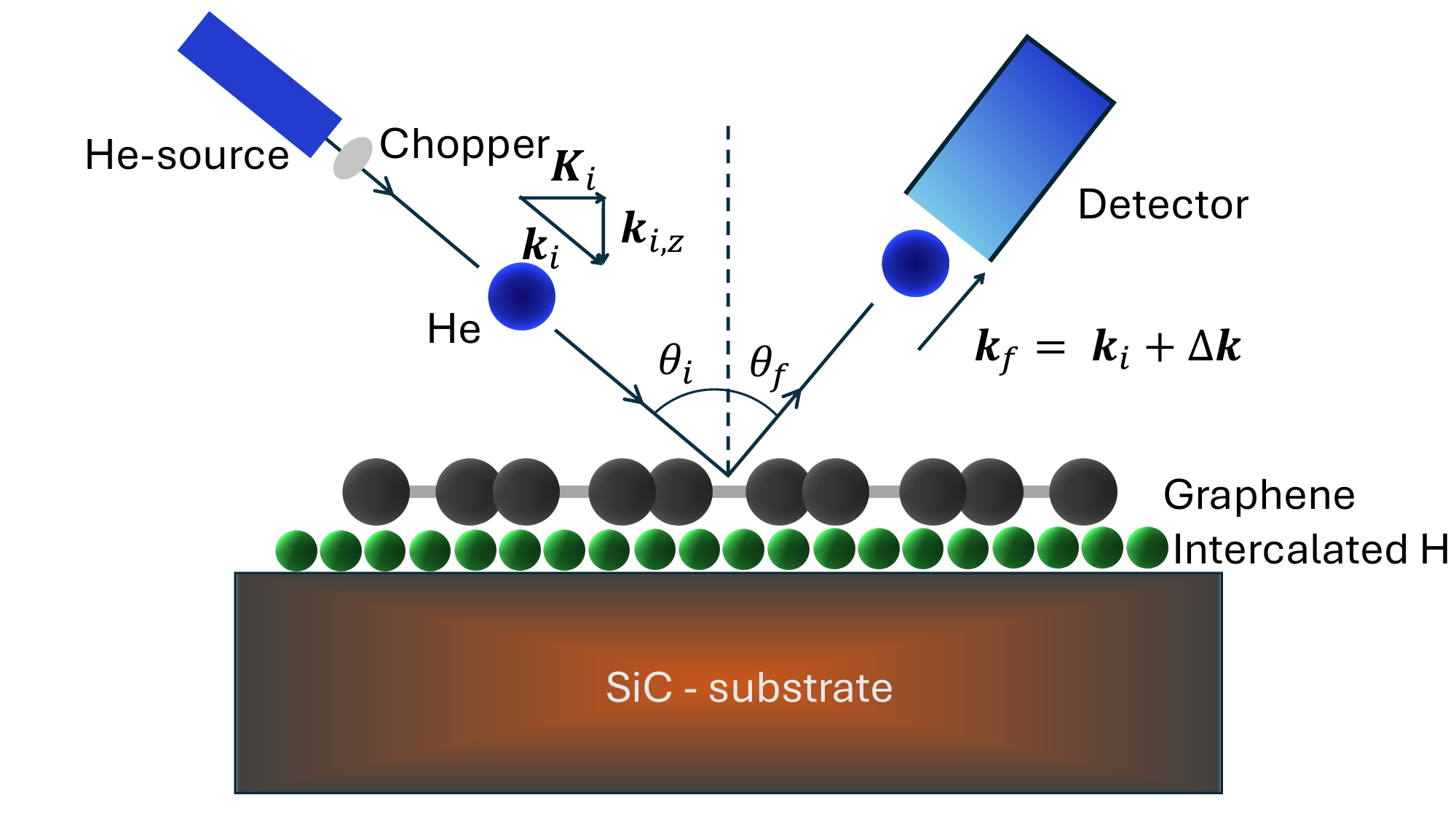}
    \caption{Illustration of the time-of-flight (TOF) helium atom scattering (HAS) experiment. Helium atoms colliding with quasi-free-standing monolayer graphene (QFMLG) supported on hydrogen intercalated silicon carbide (\ce{SiC}) are collected in the detector after collision. The incident (final) helium atom has incident (final) scattering angle $\theta_i$ ($\theta_f$) and momentum $\bm{k}_i$ ($\bm{k}_f$). The momentum vector $\bm{k}$ is split into the parallel component $\bm{K}$ and normal component $\bm{k}_z$. The source-detector angle ($\theta_i + \theta_f$) is kept fixed at 90\textdegree. From energy and momentum conservation at the scattering event, the energy $\hbar \omega_\text{ZA}$ and momentum transfer $\Delta k$ to the graphene film can be determined. }
    \label{fig:experiment}
\end{figure}
The method has since then been used to obtain $\kappa$ of 2D silica at room temperature~\cite{Buchner2018}, to explore how $\kappa$ changes with material density~\cite{Tomterud2022}, and to evaluate the temperature-dependent bending rigidity of AB stacked bilayer graphene~\cite{Eder2021}. 
 An illustration of the time-of-flight (TOF) experiment is provided in  \Figref{fig:experiment}. The TOF method allows for both diffraction and dynamical measurements in the meV range, as a room-temperature helium beam has a kinetic energy of about 63 meV~\cite{Holst2021}.

\section{Methodology}

\mnew{Here we present novel temperature-dependent measurements of $\kappa(T)$ for monolayer graphene obtained using HAS. We compare our experimental findings with a wide range of available data from the literature, as well as  temperature-dependent ZA dispersion relations computed for graphene by the means of MD simulations, based on the \texttt{Tersoff-2010}~\cite{PhysRevB.81.205441}, \texttt{Airebo}~\cite{Stuart2000}, and \texttt{lcbop}~\cite{PhysRevB.68.024107} interatomic potentials. These results complement MD~\cite{Ramirez2016} and Monte Carlo simulations~\cite{Zakharchenko2010} based on the \texttt{LCBOPII}-potential~\cite{PhysRevB.72.214102}. The collection of our results from HAS and MD simulations, together with a wide range of results derived from available literature, comprises an overview of the elastic properties of monolayer graphene, and highlights the striking experimental fact that $\kappa(T)$ increases with $T$. }

\subsection{Experimental}\label{sec2}

The measurements were performed on a quasi-free-standing monolayer graphene (QFMLG) sample prepared at the Istituto Italiano di Tecnologia (IIT) on an N-doped ($10^{18}/$cm$^3$)
6H-SiC(0001) crystal from SiCrystal GmbH. The sample was prepared using an established procedure at the institute: first the substrate was etched in a 50$\%$ Ar and H$_2$ mixture with a total pressure of 450 mbar, fluxes 500 sccm (standard cubic centimetre per minute) each, at 1200°C for 5 min to regularize the step-terrace morphology. The SiC substrate was then annealed at 1250$^{\circ}$C for 4 minutes in Ar atmosphere (780 mbar) in order to form the C-rich ($6\sqrt{3} \times 6 \sqrt{3}$)$R30^{\circ}$ reconstruction on the (0001) plane~\cite{Van_Bommel1975, PhysRevB.77.155303}. The sample was then exposed to a 50$\%$ Ar and H$_2$ mixture with a total pressure of 750 mbar (fluxes 500 sccm each) at 800°C for 20 min. This leads to atomic hydrogen intercalating at the hetero-interface between SiC(0001) and the carbon-rich surface reconstruction, thereby lifting the covalent interaction so that the carbon atoms, previously in the reconstruction, form a monolayer of graphene on top of the now H-terminated SiC(0001)~\cite{Riedl2009, Forti2011, Forti2017}. The graphene is weakly bound to the substrate: it can be peeled off the substrate in a manner similar to graphene on Cu~\cite{Liu2017}. The sample was characterized using Raman scattering and AFM at the IIT, see Supplementary information Figures S2 and S3. 
We obtain measurements of the bending rigidity of the free-standing monolayer using the phonon dispersion curve model discussed above because the sample is weakly bound. The sample temperature can be varied easily and uniformly by changing the temperature of the underlying SiC substrate.

After preparation, the sample was placed in a sealed package,
filled with ambient air from the laboratory (humidity 50$\%$, temperature 22$^\circ$C) and shipped to the University of Bergen (UiB). The sealed package was stored in the
laboratory for several months before the sample was mounted in the argon vented sample chamber of the molecular beam apparatus known as MAGIE at UiB, which was then
pumped down. 

The neutral helium beam used for sample investigation was created by a supersonic (free-jet) expansion from a source reservoir through a $10\pm 0.5~\mu$m diameter nozzle~\cite{Eder2013}. The central part of the beam was selected by a skimmer, $410\pm 2~\mu$m in diameter, placed  $11.6\pm 0.5$~mm in front of the nozzle. The sample was characterized in vacuum by helium diffraction scans and TOF spectra using a pseudorandom chopper~\cite{Koleske1992}. The background pressure during all measurements was around $1\times 10^9$~mbar or less. The incident beam spot size on the sample was around 4~mm in diameter for all measurements. 

Directly after mounting the sample and pumping it down, no diffraction pattern was observed. A bake-out was performed for 1 hour at a sample temperature $T_s = 675$~K followed by a gradual cool-down to the desired substrate temperature in 25~K steps with 60~s wait time in between. This improved the signal strongly yielding a high intensity specular peak, with a measured full width at half maximum of $(0.039 \pm0.002)$~\AA$^{-1}$ (the instrument resolution). Diffraction scans are shown in Supplementary Information Figure S1. 
This bake-out procedure was repeated between the measurements. The diffraction measurements were taken using cooled beam, $T_n = 125$~K, at a sample temperature of 400 K. The TOF measurements were taken using a beam of $p_0$ = 71 bar with beam energy $E_b = (28.9\pm 0.2)$~meV. The sample temperatures $T_s$ were (320, 370 and 400 $\pm$ 0.2)~K. An angular range of $\theta_i \in  25^{\circ}-65^{\circ}$ was investigated for each temperature, and in total 385 TOF spectra were obtained. For more details see the Supplementary Information.

\subsection{Calculational}\label{sec3}
The ZA phonon dispersion relations obtained at finite temperatures from MD simulations are largely based on the method developed by Koukaras \textit{et al.}~\cite{Koukaras2015} and post-processing of trajectories into spectral energy density is done using the \texttt{dynasor} package~\cite{Fransson2021}. The spectral energy density method is discussed in detail in, \textit{e.g.}, Ref.~\cite{Anees2015}. 
The graphene geometry is generated using ASE~\cite{HjorthLarsen2017} and MD simulations are performed using LAMMPS~\cite{LAMMPS}. The simulations were performed using a computational cell of $24 \times 24$ primitive cells, in total 1152 carbon atoms, with periodic boundary conditions.

\mnew{This work studies free-standing graphene simulated with three different interatomic potentials: The long-range carbon bond-order potential (\texttt{lcbop})~\cite{PhysRevB.68.024107}, the 2010 re-parameterization of the Tersoff potential (\texttt{Tersoff-2010})~\cite{PhysRevB.81.205441}, and the \texttt{Airebo} potential~\cite{Stuart2000}.  The calculations employ a time step of 1 fs. The system is first thermalized and relaxed for each temperature in the simulation in the $NPT$ ensemble. The thermalization lasts 0.5 ns (500,000 steps) and the relaxation lasts 5 ns (5,000,000 steps). The lattice constant of the MD trace is thereby determined as an average of the lattice constant during the relaxation in the $NPT$ ensemble. See the Supplementary Information Figure S4. The obtained lattice constant is then used in the $NVT$ simulation to obtain MD traces. The relaxed structure is assigned random velocities for all atoms, drawn from a Gaussian distribution. The MD simulation is also thermalized for 0,5 ns (500,000 steps) before running for 1 ns (1,000,000 steps).}

\mnew{Finite temperature results in the generation of thermal ripples in the graphene membrane~\cite{Deng2016}. Different initial atomic velocities, as are assigned randomly in the MD simulations, would lead to different ripples forming. Therefore, we perform} multiple (5) realisations of the MD traces at each temperature, and the phonon frequencies were obtained from the averaged spectral energy density. Examples of the phonon dispersion, as well as the spectral energy density, are shown in Supplementary Information Figures S5-S7.

\section{Results and Discussion}\label{sec4}
\subsection{Experimental results}

\begin{figure}
    \centering
    \includegraphics[width = \columnwidth]{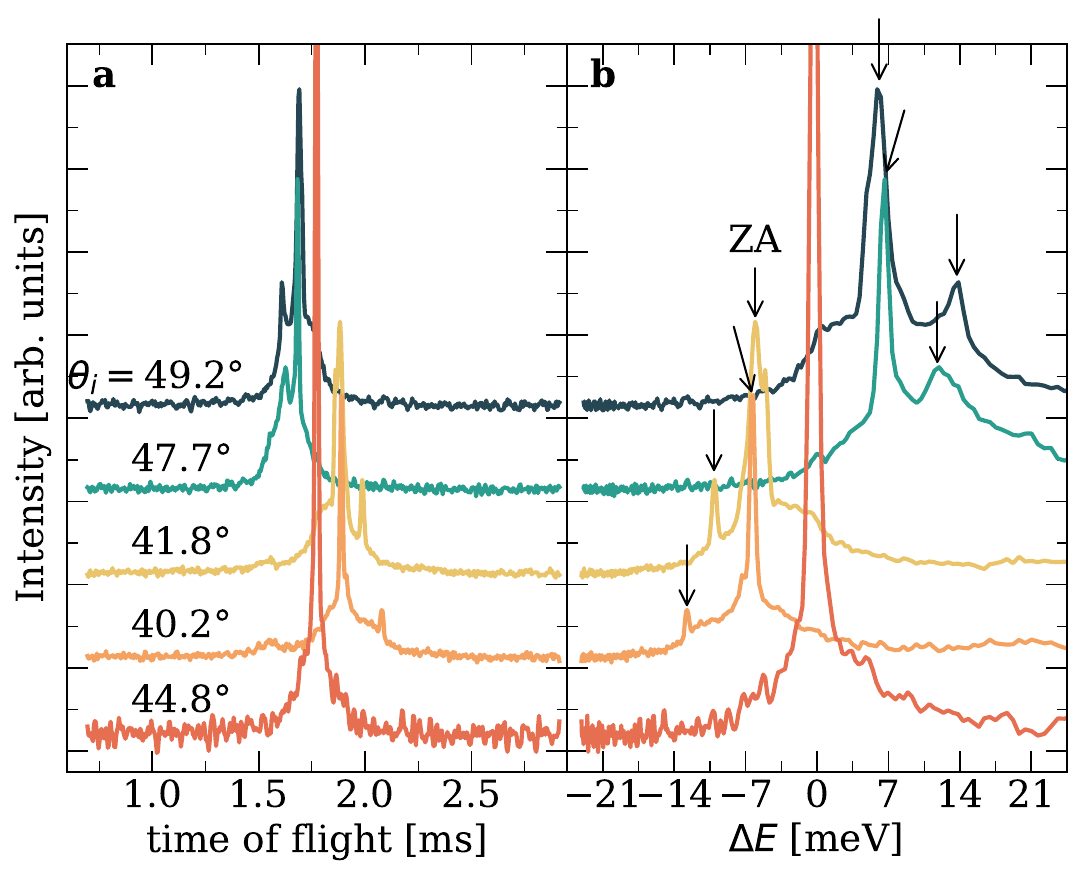}
    \caption{Time-of-flight (TOF) spectra of quasi-free-standing monolayer graphene. (a) Intensity as a function of the helium atom flight time. The incident angle $\theta_i$ for each experiment is indicated by annotations. (b) Converted data from TOF to energy exchange $\Delta E$ with the surface. Arrows identify ZA peaks in the spectra. There are no ZA excitations in the lowermost spectrum ($\theta_i=44.8^\circ$), which was taken close to specular incidence: the elastic peak is dominant in this spectrum. The spectra are vertically offset for visual clarity.}
    \label{fig:tof}
\end{figure}
\begin{figure}
    \includegraphics[width = \columnwidth]{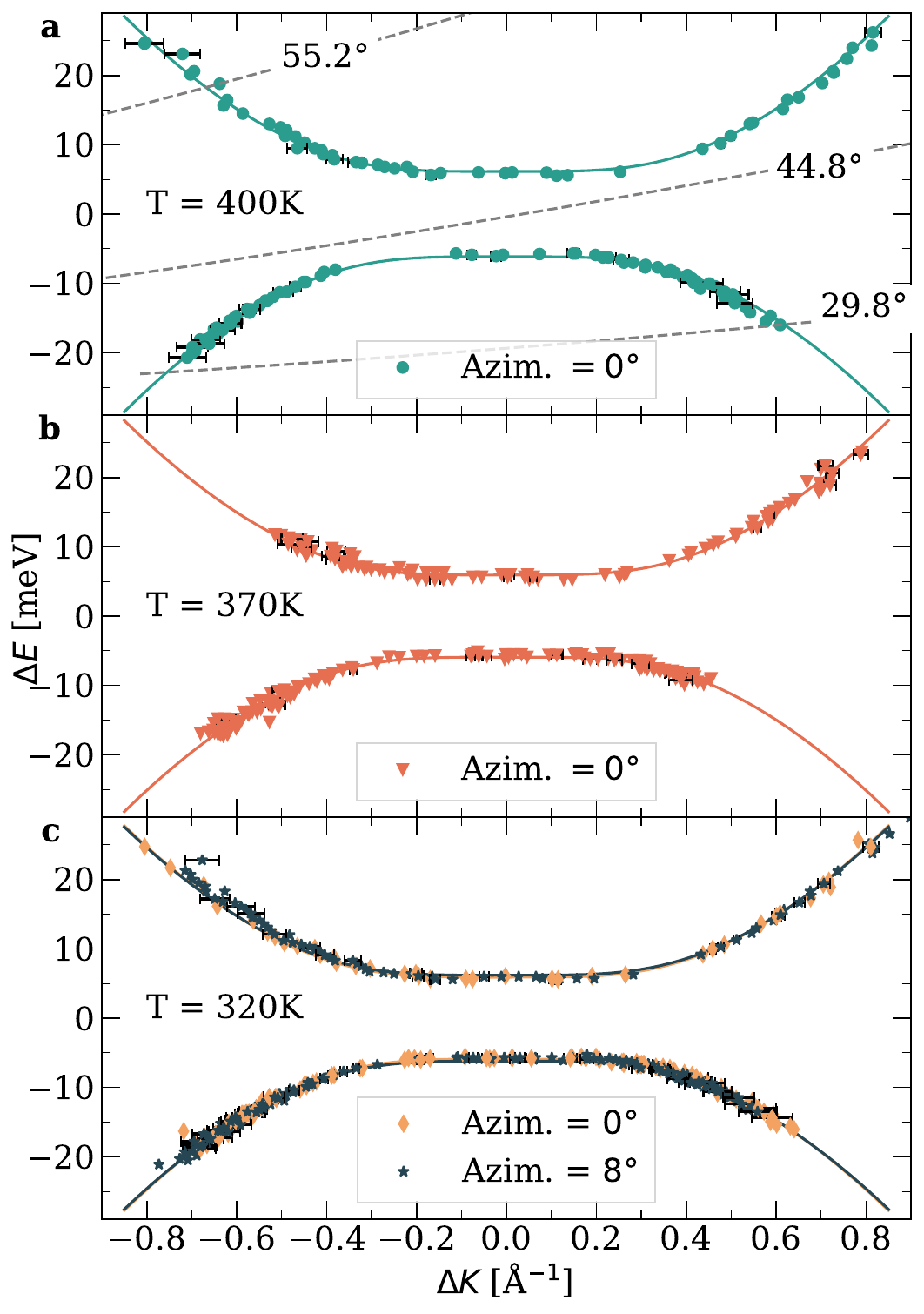}
    \caption{ZA-mode excitations of quasi-free-standing monolayer graphene measured with inelastic HAS. The modes are fitted with the dispersion relation in \Eqref{eq:two}, shown as solid coloured lines. The different panels show measurement data for different sample temperatures, as indicated. In the lowermost panel, $T=320$~K, two measurement series are shown, one of which rotated $8$\textdegree out of plane to investigate the isotropy of the ZA mode. \mnew{Panel a features scan curves for three incident angles, drawn with a dashed grey line. The scan curves show the energy exchange accessible for a single TOF experiment as a function of the parallel momentum exchange. The incident angles are indicated with inserted values along the scan curve.} For visual clarity, error bars (one standard deviation) are only drawn for 20~\% of the measurement points (randomly selected).}
    \label{fig:one}
\end{figure}

Examples of TOF spectra obtained with HAS are shown in \Figref{fig:tof}. The arrow annotations indicate the positions of ZA peaks in the spectra in \Figref{fig:tof}b. 
Identified ZA mode excitations in the HAS spectra are plotted in \Figref{fig:one}. \mnew{As indicated in \Figref{fig:tof}, each TOF spectrum can contain up to two ZA  excitations. This is explained by the scan curves drawn with dashed grey lines in \Figref{fig:one}a. The scan curve is a visualization of the energy and parallel momentum exchange of a single TOF experiment. For each intersection with the ZA dispersion, a ZA peak is visible in the TOF spectrum.} Our data examines the dispersion relation as a function of parallel momentum exchange $\Delta K$. The solid curves show fits of the dispersion relation presented in \Eqref{eq:two} to the ZA modes. The three panels correspond to different sample temperatures, with the measurement series at $T = 320$~K also depicting a measurement taken with the sample azimuth rotated $8$\textdegree, to ensure that the sample behaved isotropically in the plane as expected.
%-----------------------FIGURE-----------------------
%-----------------------END FIGURE----------------------
The ZA mode is observed to disperse in agreement with the model of \Eqref{eq:two} (continuous line). The 2D mass density of graphene used to evaluate \Eqref{eq:two} is $\rho_{\text{2D}} = 7.6 \times 10^{-7}$ kg/m$^2$, \textit{i.e.}, the mass of  two carbon atoms in the graphene unit cell with the equilibrium carbon-carbon bond length of $1.42$~\AA computed with DFT~\cite{PhysRevB.80.033407}.

\begin{figure}
    \includegraphics[width =\columnwidth]{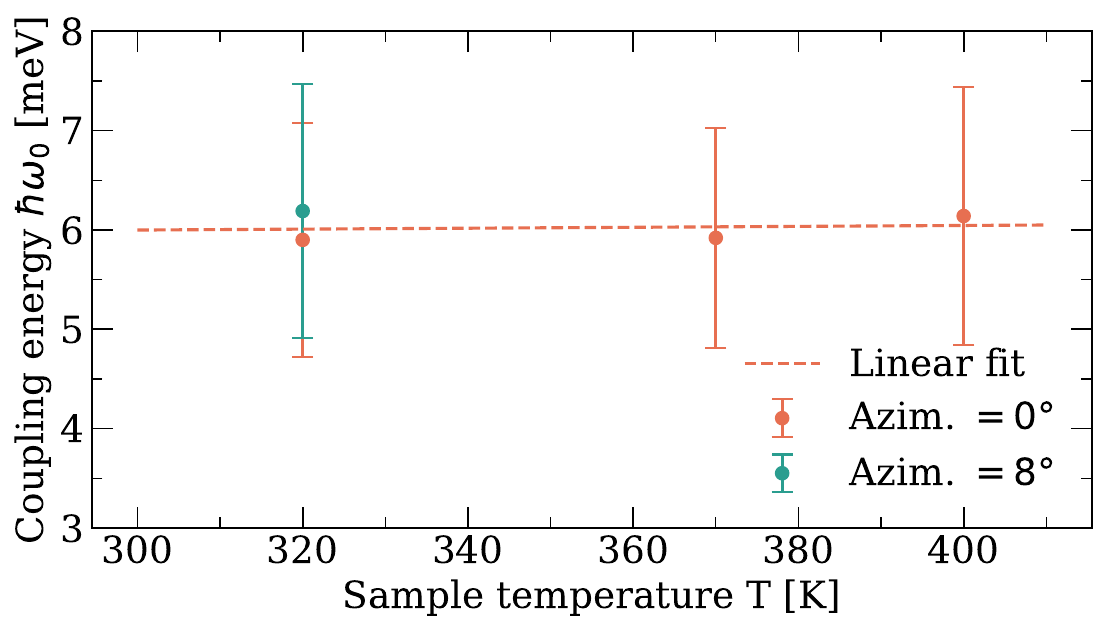}
    \caption{The extracted value of the coupling energy $\omega_0$ from fitting \Eqref{eq:two} to the experimental data in \Figref{fig:one}.  No significant change in the coupling energy is observed with increasing temperature.}
\label{fig:three}
\end{figure}

The experimental results for the coupling energy $\hbar \omega_0$ between the graphene monolayer and the SiC-substrate are presented in \Figref{fig:three}. We do not observe a significant change with $T$ in the coupling energy within the investigated temperature range. This differs from the observation made for AB stacked bilayer graphene, where the coupling energy was observed to slightly decrease with increasing temperature~\cite{Eder2021}. However, within error bars the trends of the experiments are similar. 
A summary of the experimental results for the bending rigidity and coupling energy are presented in \Tabref{tab:exp_results}.
\begin{table}
\caption{Experimental results extracted from the phonon dispersions shown in \Figref{fig:one}.}
\label{tab:exp_results}
\begin{tabular}{@{}ccccc@{}}
\toprule
 $T$ [K] & 320 & 320 & 370 & 400 \\ \midrule
 Azimuth angle & 8\textdegree & 0\textdegree & 0\textdegree & 0\textdegree \\
 $\kappa$ [eV] & $1.51 \pm 0.03$ & $1.54 \pm 0.03$ & $1.60 \pm 0.03$ & $1.63 \pm 0.03$ \\
$\hbar \omega_0$ [meV] & $6 \pm 1.5$ & $6 \pm 1.5$ & $6 \pm 1$ & $6 \pm 1.5$ \\ \bottomrule
\end{tabular}
\end{table}

\subsection{Fit to experimental data}

\begin{figure}
    \centering
    \includegraphics[width = 0.9\columnwidth]{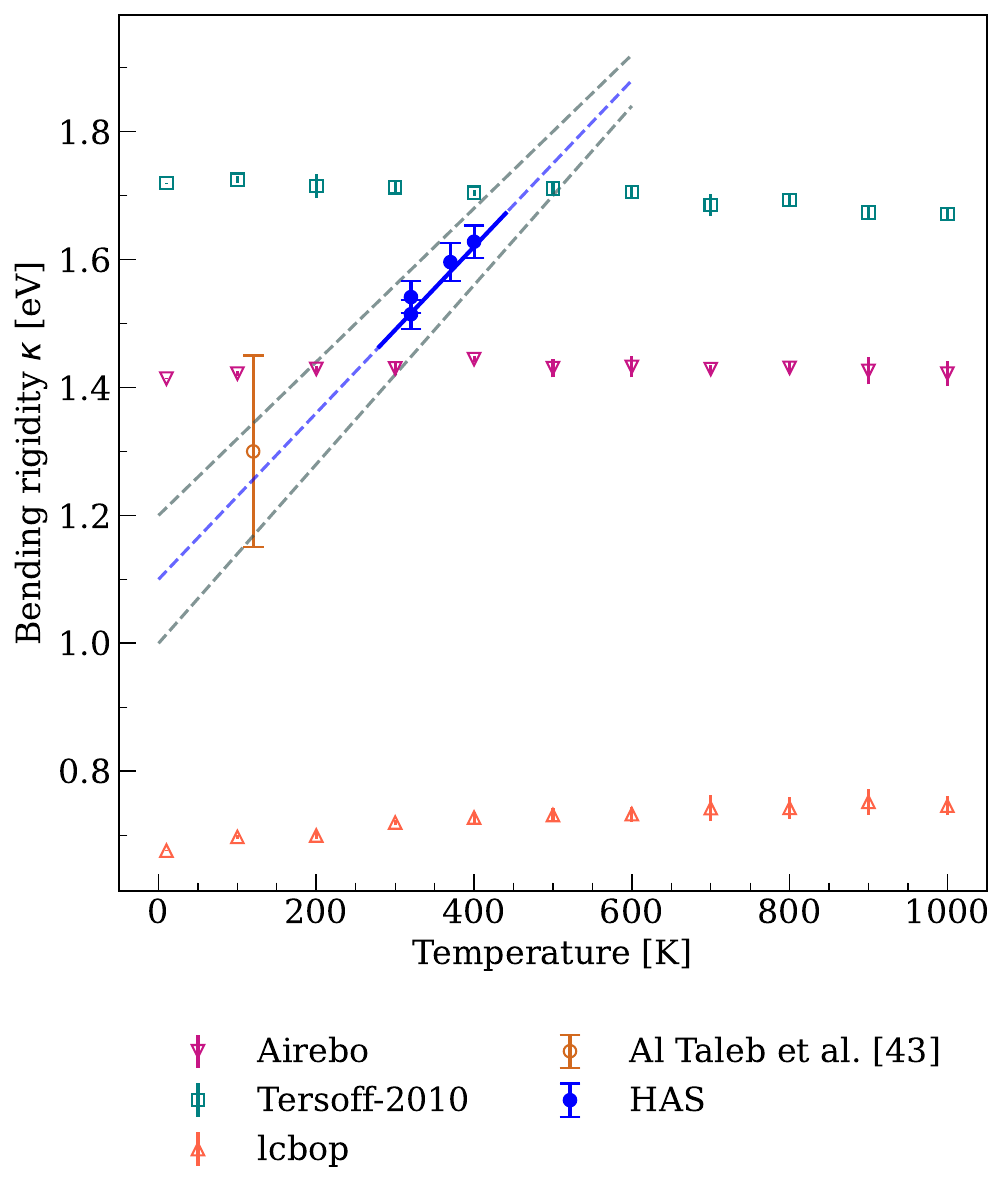}
    \caption{Calculation results of the bending rigidity of monolayer graphene, overlaid on the helium atom scattering results. The theoretically calculated bending rigidity is obtained through fits of \Eqref{eq:two} to the experimentally measured ZA dispersion. The error bars of the HAS measurements are one standard deviation of the fit. The error bars of the calculation results are one standard deviation of the bending rigidity over 5 individual MD traces.}
    \label{fig:comp_results}
\end{figure}

The temperature-dependent measurements of graphene $\kappa(T)$ are presented in \Figref{fig:comp_results} as filled circular points with error bars. As can be seen, the bending rigidity is experimentally observed to increase with temperature.
A weighted linear fit of $\kappa(T)$ based on the experimental measurements of this work is plotted with a solid blue line, and is given by
\begin{align}
\begin{split}
\kappa (T) &= (1.1 \pm 0.1)[\textrm{eV}] \\ 
& \quad + (0.0013 \pm 0.0001)[\textrm{eV/K}] \times T.
\end{split}
\label{eq:three}
\end{align}
The error bars are one standard deviation of the weighted fit. The fit is performed on a narrow temperature range, and it is not \textit{a priori} evident that a linear relation between $\kappa$ and sample temperature $T$ is correct for a larger temperature range. Therefore, we have used a solid line only in the range $T \in [280, 440]$ K. With a dashed, blue line we have extrapolated this fit as a ``best case" scenario between $T = 0$ and $600$~K. With dashed dark grey lines we have made an envelope with the standard deviations from the fit in \Eqref{eq:three}. According to this fit the bending rigidity at room temperature (300 K) is about $\kappa_{\text{fit}}(300 \text{K}) \approx 1.49$ eV.

\Figref{fig:comp_results} also includes the HAS measurement of monolayer graphene supported on copper foil by Al Taleb \textit{et al.}, mentioned in the introduction~\cite{AlTaleb2015}. This measurement was obtained at a sample temperature of 120~K and fits well, within error bars, to the linear extrapolation of the measurements series obtained for SiC-supported graphene. This measurement point is also within the envelope of our fit. There also exist HAS measurements of graphene supported on Ni(111) by Al Taleb \textit{et al.}~\cite{AlTaleb2016}. However the coupling between graphene and Ni is far stronger than the coupling between both graphene and SiC and graphene and Cu. This results in the Rayleigh mode of the Ni(111) surface being dominant in the TOF measurements for that system. 

\subsection{Comparison with our temperature-resolved MD simulations}
To gain further insight into the trend observed experimentally, we performed MD simulations to compute ZA phonon frequencies at different temperatures, and from them the corresponding temperature-dependent bending rigidity $\kappa(T)$. The results, shown in \Figref{fig:comp_results} with squares, triangles, and inverted triangles, are produced by fitting \Eqref{eq:two} to the ZA phonon dispersion obtained from each of the three potentials that were examined.  As can be seen in \Figref{fig:one}, the majority of the experimental measurements fall within the range  $\lvert\Delta K \rvert < 0.7$~\AA$^{-1}$. This is therefore taken to be the fitting range for determining $\kappa$ from the model in \Eqref{eq:two} based on the experimental measurements. \mnew{The determination of $\kappa$ using \Eqref{eq:two} is dependent on the mass density $\rho$. As the lattice constant has been determined for each potential and each simulation temperature, a temperature-dependent mass density, $\rho(T)$, is used to evaluate $\kappa$ from the MD simulation results.}

\mnew{The behaviour of $\kappa$ derived from the MD calculations is different than the one observed with HAS. The \texttt{Airebo} calculation (inverted triangles) exhibits little variation in temperature across the span of temperatures investigated. The \texttt{Tersoff-2010} calculation (squares) is observed to slightly decrease. The values obtained from the \texttt{lcbop} simulation are numerically much smaller than the other two. Nevertheless it is the only calculation observed to increase with temperature. The smaller magnitude of the bending rigidity can likely be attributed to the fact that the overall ZA dispersion is smaller in magnitude. This was also the case in the computations by Koukaras \textit{et al.}~\cite{Koukaras2015}, which can be seen by, \textit{e.g.}, comparing the ZA value at the M point. A smaller phonon frequency implies that the dispersion does not need to grow as rapidly from the $\Gamma$ point, which translates to a smaller value of $\kappa$ for the same mass density. This is exactly what we observe in the present simulations. A direct comparison of all three ZA dispersions calculated at $T = 300$ K is included in the Supplementary Information, Figure S7d. The figure shows that the \texttt{Tersoff-2010} potential has the highest ZA frequency at the M point, followed by the \texttt{Airebo}-potential and lastly the \texttt{lcbop}. The same decreasing order is also observed in the calculated $\kappa$ values.}

%The bending rigidity is observed to increase with $T$, but not with the same slope as the HAS measurements. The \texttt{Airebo} calculation (triangles) has a slightly steeper slope than the \texttt{Tersoff-2010} calculation (squares). Numerically, the \texttt{Airebo} results are closer to the experimental observations. \mnew{The values obtained from the \texttt{lcbop} simulation are numerically much smaller than the other two. Nevertheless the temperature slope is similar. The smaller magnitude of the bending rigidity can likely be attributed to the fact that the overall ZA dispersion is smaller in magnitude. This was also the case in the computations by Koukaras \textit{et al.}~\cite{Koukaras2015}, which can be seen by, \textit{e.g.}, comparing the ZA value at the $M$ point. A smaller phonon frequency implies that the dispersion does not need to grow as rapidly from the $\Gamma$ point, which translates to a smaller value of $\kappa$ for the same mass density. This is exactly what we observe in the present simulations.}

\subsection{Comparison with data from literature}
\begin{figure}
    \includegraphics[width = \columnwidth]{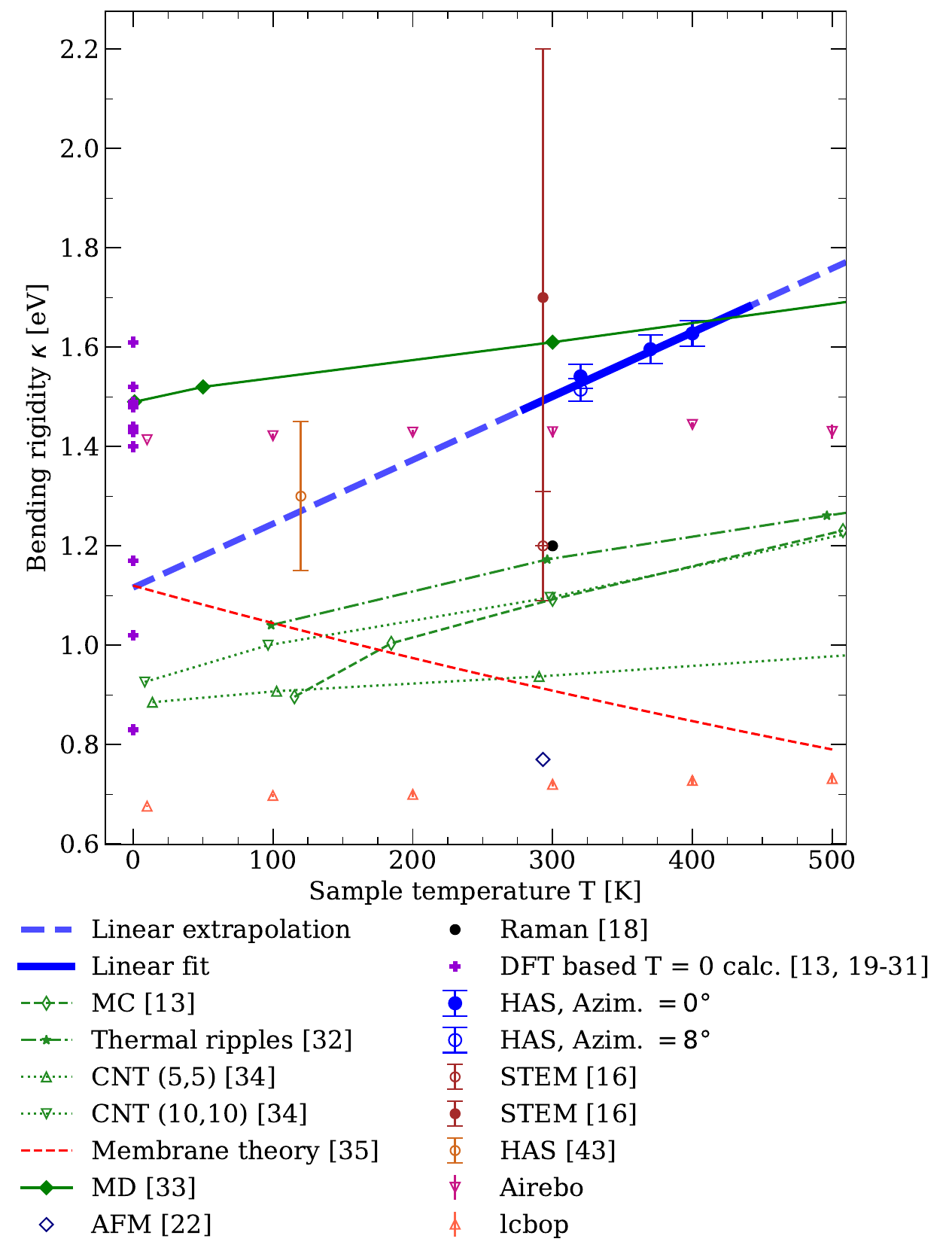}
    \caption{An overview of different results, obtained through both experiments and computations, of the bending rigidity of monolayer graphene. The present experiments are shown as blue circular points, with a linear fit of $\kappa$ as a function of sample temperature shown as a blue, solid line. The linear extrapolation to $T = 0$~K of the present experiments is in agreement with previous published work of monolayer graphene on copper foil~\cite{AlTaleb2015}, as well as MD simulations at 300 K~\cite{Ramirez2016}. Although not explicitly mentioned in the paper, it is assumed that the Raman measurements by Nicklow~\textit{et al.}~\cite{NicklowPRB} were conducted at room temperature. Among the temperature dependent results, results for carbon nanotubes (CNT) are also included~\cite{Yi-Aip}.}
    \label{fig:two}
\end{figure}

 To obtain a more complete picture of the behaviour of $\kappa(T)$, it is useful to put the HAS experiments and MD calculations into context with calculations and experiments available in the literature. This comparison is shown in \Figref{fig:two}. 
 The experimental measurements obtained with HAS are within the wide error bars reported by Han \textit{et al.}~\cite{Han2020} (brown circles), but well below the range reported by Zhao \textit{et al.}~\cite{Zhao2015} (1.8-2.7 eV, not shown). The HAS measurements are also higher than the commonly cited measurement on graphite by Nicklow \textit{et al.}~\cite{NicklowPRB} (black circle). \Figref{fig:two} also includes a range of theoretical results for $\kappa$ at zero temperature (purple symbols). The methods used include DFT~\cite{Wei2013, Sanchez-Portal-PRB, Kudin-PRB}, tight-binding DFT (DFTB)~\cite{Koskinen2010, Munoz-Diamond}, and empirical potential calculations~\cite{Tersoff-PRB, Tu-PRB, Arroyo-PRB, Lu-APD}. At finite temperature, theoretical results include elastic membrane calculations~\cite{Liu-APL} (red dashed line), calculations of the bending rigidity of carbon nanotubes~\cite{Yi-Aip} (green open triangles), Monte Carlo simulations~\cite{Zakharchenko2010} (green open diamonds) and the bending rigidity of graphene from MD simulations~\cite{Costamagna-PRB, Ramirez2016} (green stars and green filled diamonds, respectively). Both the HAS measurement presented in this work and that of Al Taleb \textit{et al.}~\cite{AlTaleb2015} sit above most theoretical predictions at finite $T$, save for the MD simulations by Ramírez \textit{et al.}, which overshoots it~\cite{Ramirez2016}.  
 
 At zero temperature there is a significant disagreement between the theoretical predictions for $\kappa$. The most important disagreement at finite temperature is whether the bending rigidity increases or decreases with increasing temperature. Reference~\cite{Liu-APL}, which disagrees with the others, is, as mentioned in the introduction, a membrane theory. This result was debated in Ref.~\cite{Zakharchenko2010}, and compared with results obtained for a liquid membrane, where the bending rigidity is known to decrease~\cite{PhysRevLett.54.1690}. Arguments have already been presented in Ref.~\cite{Zakharchenko2010} as for why the membrane approach is inappropriate, and why $\kappa$ decreases for liquid membranes and increases for crystalline materials. One interesting implication of these arguments is that for a glassy 2D material, the bending rigidity could decrease when heated. 

%-------------------BEGIN FIGURE ------------------
%-------------------END FIGURE ------------------

The temperature-dependent calculations of the \texttt{Airebo} and \texttt{Tersoff-2010} potentials shown in \Figref{fig:comp_results} have a greater value than most of the results reported in \Figref{fig:two}, but are similar in value to the MD simulation results by Ramírez \textit{et al.}~\cite{Ramirez2016}, in particular our \texttt{Airebo} calculation. Their simulations were performed with a different potential (\texttt{LCBOPII}), which likely contributes to the difference observed. The three MD calculations feature slopes that are significantly less steep than the experimental work. The results reported in Ref.~\cite{Zakharchenko2010} are also obtained with the \texttt{LCBOPII} potential, but are substantially lower than the results obtained here and those presented in Ref.~\cite{Ramirez2016}. These results are however obtained from Monte Carlo simulations, and the post-processing involves computing height correlations instead of fits to the ZA dispersion relation. The \texttt{Tersoff-2010} potential~\cite{PhysRevB.81.205441} is tailored to provide an accurate description of in-plane phonon dispersion data for graphite and the \texttt{Airebo} potential~\cite{Stuart2000} was originally designed to model chemical and intermolecular reactions in hydrocarbon systems. The long-range carbon bond-order potential (\texttt{LCBOPII}) is, among other things, meant to reproduce dissociation energy curves and improve reactive properties. 

Throughout this work we have treated the bending rigidity $\kappa$ as the proportionality constant accessible from experiments via the model in \Eqref{eq:two}. Other authors, especially those studying 2D membranes and materials theoretically, often apply the definition $\kappa \equiv \lim_{q \rightarrow 0} \omega / q^2$. This definition is applied in the recent body of work by Aseginolaza \textit{et al.}~\cite{Aseginolaza2024}, where the intrinsic properties of free-standing graphene are explored both with an atomistic model and membrane theory. Some authors have argued that the quadratic dispersion presented in \Eqref{eq:two} attains a linear term for the so-called flexural mode that makes the bending rigidity diverge as $q \rightarrow 0$~\cite{Nelson1987, PhysRevLett.69.1209}. The authors of Ref.~\cite{Aseginolaza2024} are able to go to extremely small $q$ and show that the bending rigidity of this theoretically envisioned graphene does not diverge. Within their membrane model they observe that the bending rigidity increases from $T = 0$ to $300$ K, with a slope similar to what we deduce from our \texttt{lcbop} calculations. 

The different bodies of temperature-resolved theoretical work displayed in \Figref{fig:two} exhibiting increasing behaviour with $T$ all show a similar slope, which is not as steep as the one fitted from experimental measurements. The strongest argument for a steeper slope is the agreement between the independent HAS experiments performed in this work and by Al Taleb \textit{et al.}~\cite{AlTaleb2015}. This is further strengthened by the fact that the graphene samples investigated are supported on different substrates. 

\begin{figure} 
    \centering
    \includegraphics[width = \columnwidth]{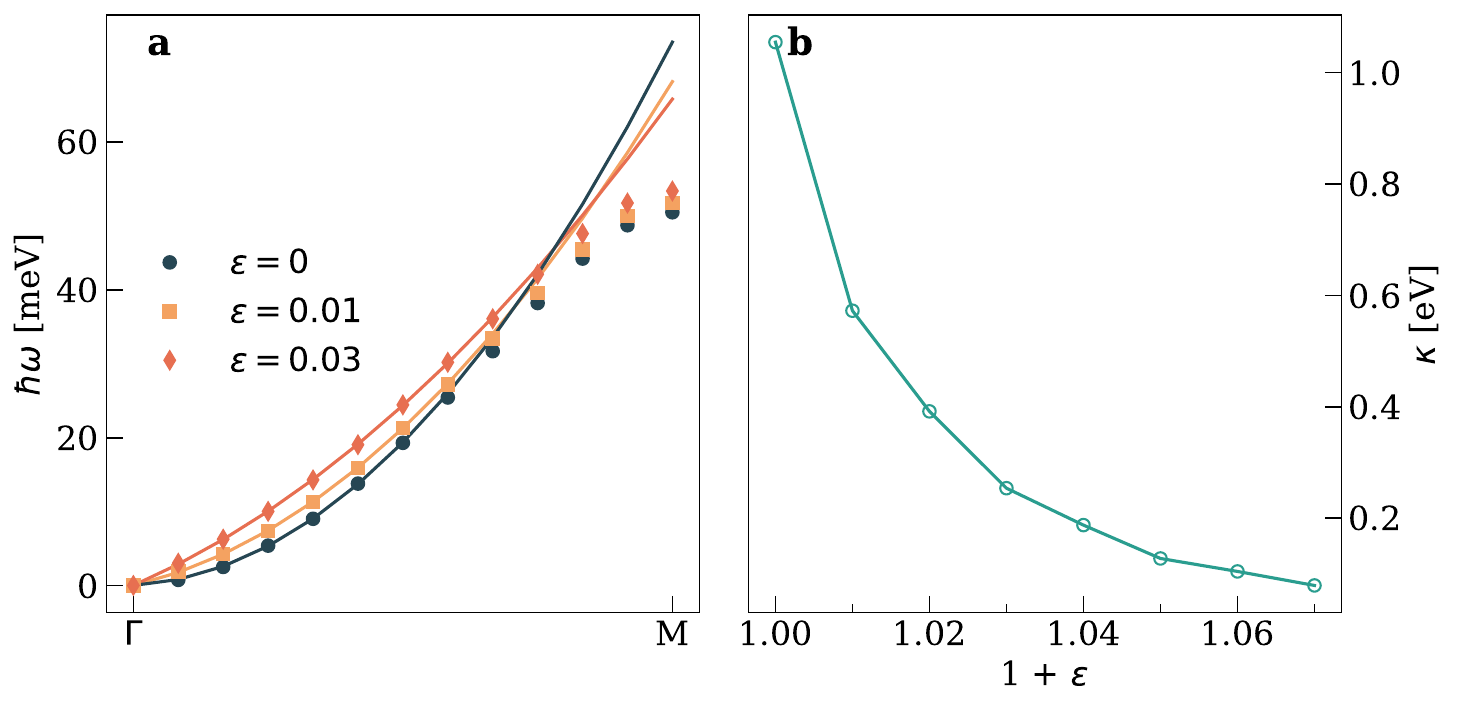}
    \caption{\mnew{Results from calculations at $T = 300$ K using the \texttt{Airebo} potential. (a) Dispersion of the ZA mode along $\Gamma$M calculated for different values of the strain parameter $\epsilon$. Solid lines are fits using the linear term model. (b) Bending rigidity of graphene as a function of the strain parameter $\epsilon$ fitted with the \Eqref{eq:za_lin} containing the linear term in the wave vector.}}
    \label{fig:strain}
\end{figure}

There is experimental evidence of strain in quasi-free-standing graphene on \ce{SiC}~\cite{Melios2016}. For the SiC-supported graphene the corrugation is only 0.02 \AA \  \cite{PhysRevLett.114.106804}. It is therefore likely that the quasi-free-standing graphene studied with HAS is under some light strain. Any strain in the system will affect the acoustic phonon modes~\cite{Aseginolaza2024}. Contrarily, all our theoretical results presented thus far are based on calculations that relax the atomic structure with the aim to remove all strain (\textit{i.e.}, targeting the self-consistent lattice constant of the method). We therefore investigated whether strain alone can be a contributing factor to the slope difference observed between experimental and computational results. \mnew{The strain calculations were performed with different time scale parameters than the main computations, see details in the Supporting Information.} \mnew{Results from calculations performed at a fixed temperature with increasing strain of the relaxed lattice constant are shown in \Figref{fig:strain}.} Fitting the dispersion relation along $\Gamma$M with the model for the ZA mode in \Eqref{eq:two} reveals that $\kappa$ increases with strain. However, the strain also linearises the dispersion. \mnew{The results in \Figref{fig:strain} are therefore shown with fits using the model with the linear term in \Eqref{eq:za_lin}.} These results show that the dispersion attains a finite linear term, which decreases the value of $\kappa$. Panel a shows the linearisation of the dispersion with increasing strain parameter $\epsilon$, and panel b shows the decrease in $\kappa$ with increasing strain, obtained in the formalism of \Eqref{eq:za_lin}. The steepest decrease of $\kappa$ with strain is in the low strain limit (less than 1\% of the lattice constant). This is surprising considering results obtained by Cao \textit{et al.}, which shows near ideal mechanical performance of graphene for much larger strain \cite{Cao2020}. It is clear mathematically that a finite linear term will decrease the value of the quadratic term. This linear term is hidden experimentally and has no influence on $\kappa$ obtained with HAS.

With HAS we are measuring the phonon dispersion over a range of $\Delta K$ out to about $0.7$~\AA$^{-1}$, with some points as far out as about $0.8$~\AA$^{-1}$. We observe that the ZA mode peaks in our spectra follow the dispersion relation in \Eqref{eq:two} up until this cutoff. However, when computing free-standing graphene phonon dispersions with MD one obtains the entire phonon spectrum, and the bending rigidity is often computed using the definition discussed above, in the limit of $q \rightarrow 0$. This part of the dispersion relation is hidden by the coupling energy in the substrate-graphene system and is not able to be probed with HAS unless one obtains a truly free-standing graphene system with $\omega_0 = 0$. Our measurements show that the ZA mode follows the dispersion relation presented in \Eqref{eq:two} at least until $0.7$~\AA$^{-1}$, which signals that obtaining the bending rigidity by only fitting close to the $\Gamma$-point could neglect contributions from further out in the Brillouin zone. Both Zakharchenko \textit{et al.}~\cite{Zakharchenko2010} and Ramírez \textit{et al.}~\cite{Ramirez2016} discuss the importance of choosing the fitting range where the slope can be approximated by the harmonic behaviour of a continuous membrane in order that the quadratic dispersion of the ZA mode be valid. The experimental data presented herein indicate that this region could include shorter wavelengths (\textit{i.e.}, larger values of $q$) than previously thought. 

\section{Conclusion}\label{sec5}
We have presented temperature-resolved HAS measurements of the ZA mode of quasi-free-standing monolayer graphene. From the phonon dispersion we extracted temperature-dependent values of the bending rigidity $\kappa$, revealing an increase with temperature.
This trend is in agreement with previously published theoretical predictions obtained using MD and Monte Carlo. We see little to no effect of temperature dependence in our own MD simulations. Although there are some theoretical results showing an increasing value of $\kappa$ with temperature, none of the theoretical predictions reproduce the experimental values within error bars over the entire temperature range. 

Future work could include more measurements on graphene over a larger temperature range. Also, measurements and calculations on more 2D materials available as weakly bound systems such as trilayer graphene, twisted bilayer graphene, bilayer silica, and other 2D materials such as h-\ce{BN}, \ce{MoS2}, and \ce{WS2}, could reveal further insights into the mechanical properties of this class of materials. In addition, bending rigidity calculations of substrate-supported 2D materials could help to better understand the difference between experimental and theoretical results. \mrevrev{Moreover, direct comparisons between the bending rigidity obtained from the ZA dispersion and the bending rigidity obtained from computing the height-height correlation function, such as in the work by Zakharchenko \textit{et al.}~\cite{Zakharchenko2010}, would be of interest.} 
 Further studies are needed to understand the influence of strain on the bending rigidity.

\clearpage
\subsection{Supplementary Information}
Supplementary information is provided with this paper.

\subsection{Acknowledgments}
B. H. acknowledges funding from the Norwegian Research Council, Grant No. 324183.
M. T. acknowledges travel support from the Norwegian Research Council, Grant No. 337339.
T. F. acknowledges support from the Spanish MCIN/AEI/10.13039/501100011033 (\mnew{Grant Nos.}~PID2020-115406GB-I00 \mnew{and PID2023-146694NB-I00}) and the European Union's Horizon 2020 (FET-Open project SPRING Grant No.~863098).
Calculations were performed at the DIPC Supercomputing Center (SCC).

\subsection{Conflict of interest}
The authors declare no conflicts of interest. 

\subsection{Availability of data and materials}
MD input scripts are available through Zenodo~\cite{tomterud_2024_dataset}.

\subsection{Authors' contributions}
S. F., D. C., and C. C. made the sample and performed Raman scattering, AFM measurements, and the associated data analysis. B. H. designed the HAS experiments. S. K. H and S. D. E. performed the HAS experiments. M. T. performed calculations and data analysis, partly as a guest at the DIPC. J. R. M. provided theoretical support. T. F. supported the calculations. M. T. and B. H. wrote the paper with input from all authors.  

\bibliography{bibliography}

\end{document}